\documentclass[12]{article}
\usepackage{fleqn}
\usepackage{graphicx}
\begin{document}
\Large
\begin{center}{\bf
The Projective Line Over the Finite Quotient Ring GF(2)[$x$]/$\langle x^{3} - x \rangle$ and
Quantum Entanglement\\
II. The Mermin ``Magic" Square/Pentagram
}
\end{center}
\vspace*{.4cm}
\begin{center}
Metod Saniga,$^{\dag}$ Michel Planat$^{\ddag}$  and
 Milan Minarovjech$^{\dag}$
\end{center}
\vspace*{.0cm} \normalsize
\begin{center}
$^{\dag}$Astronomical Institute, Slovak Academy of Sciences\\
SK-05960 Tatransk\' a Lomnica, Slovak Republic\\
(msaniga@astro.sk, milanmin@astro.sk)

\vspace*{.1cm}
 and

\vspace*{-.1cm}

\vspace*{.2cm}
 $^{\ddag}$Institut FEMTO-ST, CNRS, D\' epartement LPMO, 32 Avenue de
l'Observatoire\\ F-25044 Besan\c con, France\\
(planat@lpmo.edu)
\end{center}

\vspace*{-.3cm} \noindent \hrulefill

\vspace*{.0cm} \noindent {\bf Abstract}

\noindent
In 1993, Mermin (Rev. Mod. Phys. 65, 803--815) gave lucid and strikingly simple proofs of the Bell-Kochen-Specker
(BKS) theorem in Hilbert spaces of dimensions four and eight by making use of what has since been referred to as the Mermin(-Peres)
``magic square" and the Mermin pentagram, respectively. The former is a $3 \times 3$ array of nine observables
commuting pairwise in each row and column and arranged so that their product properties contradict those of the assigned eigenvalues.
The latter is a set of ten observables arranged
in five groups of four lying along five edges of the pentagram and characterized by similar contradiction.
An interesting one-to-one correspondence between the operators of the Mermin-Peres square and the points of the projective
line over the product ring ${\rm GF}(2) \otimes \rm{GF}(2)$ is established. Under this mapping, the concept ``mutually commuting"  translates
into ``mutually distant" and the distinguishing character of the
third column's observables has its counterpart in the distinguished properties of the coordinates of the corresponding
points, whose entries are  both either zero-divisors,  or units.
The ten operators of the Mermin pentagram answer to
a specific subset of points of the line over GF(2)[$x$]/$\langle x^{3} - x \rangle$. The situation here is, however, more intricate as
there are two different configurations that seem to serve equally well our purpose.
The first one comprises the three distinguished points of the (sub)line over GF(2), their three ``Jacobson" counterparts
and the four points whose both coordinates are zero-divisors; the other features the neighbourhood of the
point ($1, 0$) (or, equivalently, that of ($0, 1$)). Some other ring lines that might be relevant for BKS proofs in
higher dimensions are also mentioned.
\\



\hspace*{2.cm}Quantum Entanglement

\vspace*{-.2cm} \noindent \hrulefill

\section{Introduction}
In Part I of our paper \cite{sp}, after highlighting the fundamental properties of commutative rings with unity,  we introduced an important
algebraic geometrical concept, namely that of a projective line defined over a (finite) {\it ring}. This concept was then illustrated in detail
on the structure of two
rather elementary kinds of projective ring line; the line defined over the factor ring GF(2)[$x$]/$\langle x^{3} - x \rangle$ and that over the
elementary product
ring ${\rm GF}(2) \otimes \rm{GF}(2)$.
Both the lines feature finite number of points, eighteen and nine respectively, shown to form three distinguished groups in term of the so-called
neighbour
and/or distant relation and endowed with a number of interesting properties. In this part we aim at demonstrating that these two remarkable
ring geometries
can be employed in mimicking the structure of the Mermin(-Peres) ``magic square" and the Mermin pentagram---the two essential ingredients
in one of the
simplest proofs of the Bell-Kochen-Specker (BKS) theorem furnished up to date \cite{mer1},\cite{mer2}.

Some five years ago, Aravind \cite{ar1} pointed out that the 24
quantum states (or rays) employed by Peres in \cite{per} to prove
the BKS theorem are intimately linked with Reye's configuration of
twelve points and sixteen lines in the classical projective space,
i.e. space defined over a {\it field}. The dodecahedron of Zimba
and Penrose \cite{zp} is another configuration with many
interesting classical projective properties which turned out to be
of relevance for BKS proofs. The present paper may thus be
regarded as a natural, qualitative extension and/or generalization
of the spirit of Aravind's and Zimba-Penrose's geometrical
reasoning into the domain of more abstract projective geometries,
where fields are replaced by rings.

\section{The Bell-Kochen-Specker Theorem and the Two\\ Mermin's Configurations}
Quantum mechanics imposes, in general, only statistical
restrictions on the results of measurements and so one is
naturally tempted to assume that it is an incomplete theory, being
the result of a more complete description in terms of so-called
hidden variables, or ``elements of reality" [7]. The
Bell-Kochen-Specker theorem \cite{bell},\cite{ks} renders such a
description impossible. It shows that any hidden-variable theory
in the Hilbert spaces of dimensions three and higher must be
contextual, i.e. relying not only on hidden states in the quantum
system under study, but also on those in the measuring devices.
There have been a large number of proofs of this theorem,
differing from each other in philosophy and a degree of
technicalities involved, but those given by Mermin in
\cite{mer1},\cite{mer2} for dimensions four and eight stand out as
most straightforward and succinct ever furnished. Here, we shall
not be interested in these proofs themselves, but focus merely on
(the properties of) two remarkable configurations of observables
which play a key/decisive role in them.

\begin{figure}[b]
\begin{center}
\begin{tabular}{ccc}
$\sigma_{x}^{1}$ & $\sigma_{x}^{2}$ & $\sigma_{x}^{1} \sigma_{x}^{2}$ \\
&& \\
$\sigma_{y}^{2}$ & $\sigma_{y}^{1}$ & $\sigma_{y}^{1} \sigma_{y}^{2}$ \\
&& \\
~$\sigma_{x}^{1} \sigma_{y}^{2}$~ & ~$\sigma_{x}^{2} \sigma_{y}^{1}$~ & ~$\sigma_{z}^{1} \sigma_{z}^{2}$~
\end{tabular}
\end{center}
\caption{The Mermin(-Peres) ``magic square" [2]. This configuration is unique up to a transposition of its rows and/or columns.}
\end{figure}

Let us start in four dimensions and, so, with the configuration usually referred to as the Mermin(-Peres) ``magic square" (see \cite{ar2} for statements
of credits).
This configuration, depicted in Fig.\,1,  represents a $3 \times 3$ array of nine observables for a system of two-qubits, with the superscripts on the
operators
referring
to the qubits and with $\sigma_{x}$, $\sigma_{y}$, $\sigma_{z}$ denoting the Pauli matrices,
\begin{eqnarray}
\sigma_{x} =
\left( \begin{array}{cc}
0 & 1 \\
1 & 0
\end{array}
\right),~~
\sigma_{y} =
\left( \begin{array}{cr}
0 & -i \\
i & 0
\end{array}
\right),~~
\sigma_{z} =
\left( \begin{array}{cr}
1 & 0 \\
0 & -1
\end{array}
\right).
\end{eqnarray}
One can easily verify that each row/column features three pairwise
commuting observables and that the product of the three operators
in each row and each of the first two columns is $+I$ (the
identity matrix), but the product of those in the {\it third}
column is $-I$. Now, each of the operators in the square can be
assigned an eigenvalue $\pm$1, and since these eigenvalues must
obey the same identities as the operators themselves, their
product in each row and the first two columns must be +1, whilst
in the last column $-1$. This is, however, impossible, since
according to the rows the parity is even, yet according to the
columns, odd; hence, the name ``magic square."
Going to eight dimensions, Mermin [2] considers three
qubits instead of two, and employs the ten relevant observables
located at the vertices of a pentagram, as shown in Fig.\,2.  The
four observables in each of the five edges of the pentagram are
mutually commuting, and their product on every edge is $+I$,
except for the {\it horizontal} one, where it is $-I$. As the same
identities must also be satisfied by the eigenvalues assigned to
the observables, such a configuration cannot exist because each
observable, and so its eigenvalue, is shared by two edges.

\begin{figure}[t]
\begin{center}
\begin{tabular}{rcccl}
&&&&\\
&& $\sigma_{y}^{1}$ & &  \\
&&&& \\
&&&& \\
$\sigma_{x}^{1} \sigma_{x}^{2} \sigma_{x}^{3}$~~ & $\sigma_{y}^{1} \sigma_{y}^{2} \sigma_{x}^{3}$& &
$\sigma_{y}^{1} \sigma_{x}^{2} \sigma_{y}^{3}$ & ~~$\sigma_{x}^{1} \sigma_{y}^{2} \sigma_{y}^{3}$ \\
&& &&\\
&&&& \\
&$\sigma_{x}^{3}$~~ && ~~$\sigma_{y}^{3}$ & \\
&& &&\\
&& $\sigma_{x}^{1}$ & &  \\
&& &&\\
&&&& \\
$\sigma_{y}^{2}$ &&&& $\sigma_{x}^{2}$ \\
&&&&\\
\end{tabular}
\end{center}
\caption{The Mermin pentagram [2]. The symbols are the same as in the previous figure, with the superscripts now referring to three distinct qubits.}
\end{figure}

The two configurations just introduced are very similar to each other as they both consist of a finite number of sets of pairwise commuting observables
of the same cardinality, where one of the sets stands on a qualitatively different footing than the rest of them. A natural question emerges: Can one find
any other algebraic geometrical configurations behaving in a similar way? The answer is yes; and to justify this answer, we only need to return
to [1; Sec.\,4].

\section{The Mermin(-Peres) Square and the Projective Line Over GF(2)$\otimes$GF(2)}
We shall deal first with a ring geometrical analogue of Mermin's
square, which turns out to be nothing but
$P\widetilde{R}_{\clubsuit}(1)$ --- the projective line defined
over ${\rm GF}(2) \otimes \rm{GF}(2)$ \cite{sp}. The nine points
of this line [1; Eqs.\,(20)--(22)] can be arranged into a $3
\times 3$ array as shown in Fig.\,3. This array has an important
property that all the points in the same row and/or column are
pairwise {\it distant}. Moreover, a closer look at Fig.\,3 reveals
that one triple of mutually distant points, that located in the
third column, differs from all the others in having both the
entries in the coordinates of all the three points of the same
character, namely either zero-divisors (the points $(x, x+1)$ and
$(x+1, x)$), or units (the point $(1, 1)$). After comparing
Fig.\,1 with Fig.\,3, and identifying, in an obvious way, the
observables of the Mermin(-Peres) square with the points of
$P\widetilde{R}_{\clubsuit}(1)$, one immediately sees that the
concept {\it mutually commuting}  translates ring geometrically
into {\it mutually distant} and that the ``peculiar" character of
the third column's observables has its geometrical counterpart in
the above-mentioned distinguishing properties of the coordinates
of the corresponding points. It is, however, worth pointing out
that it is also the third row's observables that get a piece of
recognition in our picture, for they correspond to the points
which represent nothing but the embedding in
$P\widetilde{R}_{\clubsuit}(1)$ of the {\it ordinary} projective
line over GF(2), $PG(1, 2)$.
\begin{figure}[bh]
\begin{center}
\begin{tabular}{ccc}
~$(x+1, 1)$~ & $(1, x)$ & ~$(x, x+1)$~ \\
&& \\
$(x, 1)$ & ~$(1, x+1)$~ & $(x+1, x)$ \\
&& \\
$(1, 0)$ & $(0, 1)$ & $(1, 1)$
\end{tabular}
\end{center}
\caption{An arrangement of the points of $P\widetilde{R}_{\clubsuit}(1)$ into a square array in such a way that any two points in each row/column are
distant.}
\end{figure}

\section{The Mermin Pentagram and the Projective Line Over GF(2)[$x$]/$\langle x^{3} - x \rangle$}
The structure of the Mermin pentagram is obviously more intricate and complex than that of the square, and these properties must obviously
carry over onto its ring geometrical sibling(s). The relevant ring geometry is now that of $PR_{\clubsuit}(1)$, the projective line defined over
the finite factor ring $R_{\clubsuit} \equiv$
GF(2)[$x$]/$\langle x^{3} - x \rangle$ \cite{sp}.
In fact, we have here  two different ten-point configurations that seem to serve equally well our purpose.
The first one comprises the point $(1, 0)$ (or, equivalently, $(0, 1)$) and the nine points of its neighbourhood [1; Eq.\,(15)], arranged
as shown in Fig.\,4.
The other one features the three distinguished points of the (sub)line over GF(2), viz. $(1, 0)$, $(0, 1)$ and $(1, 1)$,
their three ``Jacobson" counterparts, $(1, x^{2}+x)$, $(x^{2}+x, 1)$ and $(1, x^{2}+x+1)$, respectively,
and the four points whose both coordinates are zero-divisors, and arranged as depicted in Fig.\,5.
\begin{figure}[ht]
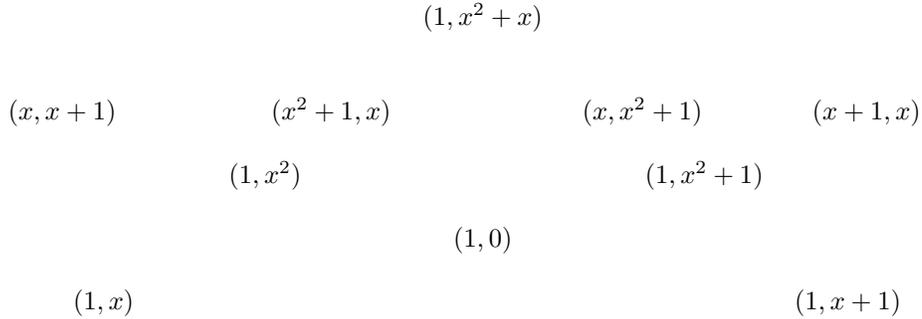

\begin{center}
\begin{tabular}{rcccl}
&&&& \\
&& $(1, x^{2} + x)$ & &  \\
&&&& \\
&&&& \\
$(x, x+1)$~~ & ~~~~~~~~~~~~$(x^{2}+1, x)$& &
$(x, x^{2}+1)$~~~~~~ & ~~$(x+1, x)$ \\
&& &&\\
&$(1, x^{2})$~~~ && ~~~~~~~~$(1, x^{2}+1)$ & \\
&& &&\\
&& $(1, 0)$ & &  \\
&& &&\\
$(1, x)$ &&&& $(1, x+1)$
\end{tabular}
\end{center}
\caption{A pentagram-forming subset of ten points of $PR_{\clubsuit}(1)$, consisting of the point $(1, 0)$
and its neighbourhood.
The uppermost vertex
of the pentagram is the ``Jacobson" point of the neighbourhood \cite{sp}. }
\end{figure}
\begin{figure}[ht]
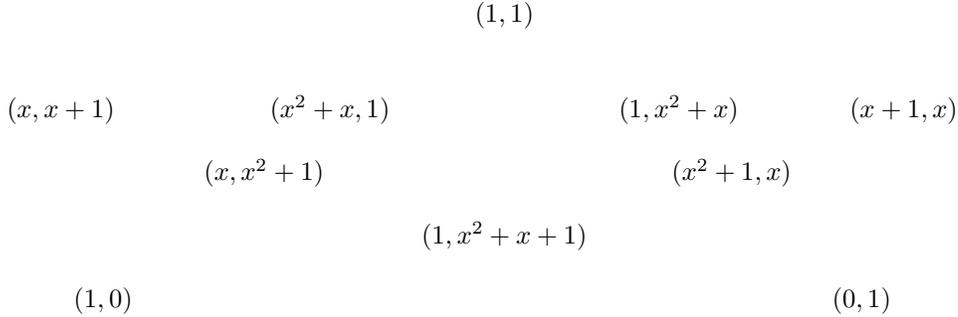

\begin{center}
\begin{tabular}{rcccl}
&&&& \\
&& $(1, 1)$ & &  \\
&&&& \\
&&&& \\
$(x, x+1)$~~ & ~~~~~~~~~~~~$(x^{2}+x, 1)$& &
$(1, x^{2}+x)$~~~~~~~ & ~~$(x+1, x)$ \\
&& &&\\
&$(x, x^{2}+1)$~~~ && ~~~~~$(x^{2}+1, x)$ & \\
&& &&\\

&& $(1, x^{2}+x+1)$ & &  \\
&& &&\\
$(1, 0)$ &&&& $(0, 1)$
\end{tabular}
\end{center}
\caption{A ten-point subset of  $PR_{\clubsuit}(1)$ made of three points of $PG(1, 2)$,
their three ``Jacobson" counterparts, and the four points whose both coordinates are zero-divisors.}
\end{figure}

In the former case, even a passing look at Fig.\,4 reveals that the distinct character of the horizonal edge is due to the sole zero-divisor entries in
the coordinates of all the four points. In the latter case, the ``prominent" character of the horizontal edge is not so readily discernible and one has
to invoke the neighbour/distant relation [1] between the relevant points to spot it. We find, in particular, that the quadruples of points at each of the
remaining edges have the property that one of the points is distant to each of the remaining three; as per the top-to-bottom-right/left edges it is the
point $(1, 1)$, whereas for the right/left-to-bottom-left/right ones this role is played by the point $(1, x^{2} + x + 1)$; there exists, however, {\it no}
such point for the horizontal edge.

Is there any means of discriminating between the two
configurations? An affirmative answer is provided by the
ring-induced homomorphism from $PR_{\clubsuit}(1)$ into
$P\widetilde{R}_{\clubsuit}(1)$ [1; Eq.\,(23)]. Under this
homomorphism, the four ``horizontal" points of the first
configuration (Fig.\,4) map into only two distinct points, namely
the $(x, x+1)$ and $(x+1, x)$ ones, whilst for the second
configuration (Fig.\,5) we get four distinct points --- the
above-given two points and the points $(1, 0)$ and $(0, 1)$.
Hence, the ``neighbourhood-generated" analogue of the Mermin
pentagram seems to be more appealing, for its horizontal edge is
homomorphic to (a portion of) the distinguished, third column of
the ring geometric analogue of the Mermin(-Peres) magic square
(Fig.\,3) and, as a whole, it ``condenses" into the set of all the four points
distant to $(1, 1)$.

\section{Zero-Divisors and Quantum (Entanglement)?}

In order to fully appreciate the meaning of our {\it ring} geometrical analogues of the Mermin's ``magic" configurations, we shall show
that they cannot be reproduced by any classical, i.e. field projective lines. We shall examine the square case only, as the
procedure can readily be extended to the pentagram case.

To this end in view, we simply notice that only four distinct marks, viz. 0, 1, $x$  and $x+1$, appear in Fig.\,3; these are, of course, the elements
of the ring $\widetilde{R}_{\clubsuit}$ [1; Eq.\,(7)]. If we consider the projective line over the field featuring the same number of elements,
GF(4) $\cong$ GF(2)[$x$]/$\langle x^{2} + x +1 \rangle$ \cite{h} then,
repeating the strategy and reasoning of [1; Sect\,4], we find that such a line features {\it only five} points, namely $(1, 0)$, $(1, 1)$, $(1, x)$, $(1, x+1)$
and $(0, 1)$,
because the elements/marks $x$ and $x+1$
represent now {\it units}.\footnote{Let us recall (see, e.g., \cite{h}) that a field does not contain any zero-divisor --- except for the trivial one (zero).}
In order to get the required number of points, we have to take the line defined over GF(8) $\cong$ GF(2)[$x$]/$\langle x^{3} + x +1 \rangle$;
the price to be paid for this move is, however, introducing additional, superfluous marks --- those featuring second powers of $x$ --- into our
scheme.

This illustration makes thus explicit what all the preceding
discussions seem to evince, namely that it is thanks to the
presence of {\it zero-divisors} in our approach that we are able
to shed some light on the {\it quantum} intricacies embodied in
the two Mermin's configurations. This claim can be made
substantially stronger by the following observations. We can
rephrase the ``magic" of the Mermin(-Peres) square (Fig.\,1) by
noticing that the three rows have joint (mutually unbiased to each
other) orthogonal bases of {\it un}entangled eigenstates and the same
holds for the first two columns; on the contrary, the operators in
the {\it third column} share a base of {\it maximally entangled}
states. And looking at Fig.\,3, we see that it is the {\it third
column} of its geometrical counterpart where the presence of
zero-divisors is {\it most pronounced}. The same applies to the
Mermin pentagram (Fig.\,3), with the horizontal edge's operators
sharing a base of {\it maximally entangled} states, and its
``neighbourhood-induced" analogue (Fig.\,4), with the
zero-divisor-dominated horizontal edge. 

\section{Conclusion}

We have drawn a remarkable, though at this stage rather subtle,
analogy between the structure of the two ``magic" operator-valued
configurations employed by Mermin in [2] to prove the BKS theorem
in dimensions four and eight and the point sets represented by the
projective ring line defined over  ${\rm GF}(2) \otimes
\rm{GF}(2)$ and sub-configurations of the line defined over
GF(2)[$x$]/$\langle x^{3} - x \rangle$. A cornerstone algebraic
geometrical concept of this correspondence is that of {\it
zero-divisors} and its closest ally, the {\it neighbour/distant}
relation \cite{sp}, which may well turn out to lead to a deeper
understanding of quantum entanglement. In order to test this
hypothesis, it will necessitate examining more general kinds of projective
ring line, in particular those defined over ${\rm GF(q)} \otimes
\ldots \otimes \rm{GF(q)}$ and/or GF(q)[$x$]/$\langle x^{s} - x
\rangle$, with $q > 2$ and $s > 3$, and see whether their
properties are indeed relevant for proofs of the BKS theorem in
higher-dimensional  Hilbert spaces.

As a final note, we would like to stress that our simple ring geometries may turn out to be just the right starting point to a more systematic, 
{\it geometrically}-oriented modelling of entangled quantum states/systems. In this sense our current findings, supplemented by the role of other finite
ring geometries, namely those of Hjelmslev, were found to play in addressing the properties of mutually unbiased bases
\cite{sp2},\cite{sp3}, lend some support to the ``Relational Block World" view of quantum mechanics, recently proposed and advocated 
by Stuckey {\it et al.} \cite{ssc}, which is a novel paradigm resting uniquely on non-dynamical, {\it geometric} explanation of quantum phenomena.
\\ \\ \\
\noindent
\Large
{\bf Acknowledgements}
\normalsize

\vspace*{.2cm}
\noindent
This work was supported, in part, by the Science and Technology Assistance Agency (Slovak Republic) under the
contract $\#$ APVT--51--012704, the VEGA project $\#$ 2/6070/26 (Slovak Republic) and  the ECO-NET project $\#$ 12651NJ
``Geometries Over Finite Rings and the Properties of Mutually Unbiased Bases" (France).

\end{document}